\documentclass[aps,prb,twocolumn,prb,showpacs]{revtex4}
\usepackage{graphicx}
\usepackage{amsmath}

\begin{document}

\pagestyle{plain}

\title{Freezing and melting equations for the $n$-6 Lennard-Jones systems}

\author{Sergey A. Khrapak$^{1,2,3}$ and Ning Ning${^1}$}

\affiliation{${^1}$Aix-Marseille-Universit\'{e}, CNRS, Laboratoire PIIM, UMR 7345, 13397 Marseille cedex 20, France; \\${^2}$Forschungsgruppe Komplexe Plasmen, Deutsches Zentrum f\"{u}r Luft- und Raumfahrt (DLR),
Oberpfaffenhofen, Germany; \\${^3}$Joint Institute for High Temperatures, Russian Academy of Sciences, Moscow, Russia}

\date{\today}

\begin{abstract}
We generalize previous approach of Khrapak and Morfill [J. Chem. Phys. {\bf 134}, 094108 (2011)] to construct simple and sufficiently accurate freezing and melting equations for the conventional Lennard-Jones (LJ) system to $n$-6 LJ systems, using the accurate results for the triple points of these systems published by Sousa {\it et al.} [J. Chem. Phys. {\bf 136}, 174502 (2012)].   
\end{abstract}

\maketitle

It was observed in Ref.~\onlinecite{KM2011} that several phenomenological approaches~\cite{Rosenfeld1,Rosenfeld2,PRB2010}  
to locate the melting/freezing transition in simple systems of interacting particles
predict the same functional dependence of the temperature on liquid and solid densities at coexistence, when applied to the conventional Lennard-Jones (LJ) potential.   
Specifically, the freezing and melting curves in the ($\rho$, $T$) plane can be well described by a simple analytic form 
\begin{equation}\label{original}
T_{{\rm L, S}}={\mathcal A}_{{\rm L, S}}\rho^4-{\mathcal B}_{{\rm L, S}}\rho^2,
\end{equation}
where the subscripts ${\rm L}$ and ${\rm S}$ correspond to liquid and solid phases, respectively. Here the reduced temperature $T$ and density $\rho$ are in units of the characteristic energy scale, $\epsilon$, and volume scale, $\sigma^3$, of the LJ interaction potential, $U_{\rm LJ}(r)= 4\epsilon\left[(\sigma/r)^{12}-(\sigma/r)^{6} \right]$.    
More recently, it has been argued that the functional form of equation (\ref{original})
also emerges naturally as a consequence of the isomorph theory and the prediction that the liquid-solid coexistence line is itself an isomorph.~\cite{IngeJCP2012,PRX2012,DyrePRE2013,Arxiv2016}  
The consistent procedure to determine the coefficients ${\mathcal A}$ and ${\mathcal B}$ in Eq.~(\ref{original}) was also proposed in Ref.~\onlinecite{KM2011}. The values ${\mathcal A}_{\rm L,S}$ were obtained from the high-temperature limit, governed by the inverse twelfth power repulsive potential. The coefficients ${\mathcal B}_{\rm L,S}$ were then determined from the triple point parameters of the LJ fluid. This procedure (different from that employed in the isomorph approach~\cite{Arxiv2016}) produced freezing and melting equations which are essentially exact in the high-temperature limit and in the vicinity of the triple point, and show reasonably good agreement with existing numerical simulation data in the intermediate region.~\cite{KM2011}

It was also recognized in Ref.~\onlinecite{KM2011} that freezing and melting equations similar to that of Eq.~(\ref{original}) can be easily constructed for a family of $n$-6 LJ systems. This was not done, however, because of the lack of reliable data for the triple point parameters of these systems. At present this gap is filled. Sousa {\it et al}.~\cite{Sousa2012} reported detailed determination of the solid-fluid coexistence for $n$-6 LJ systems from precise free energy consideration. The triple point parameters are tabulated in Table V of their paper. These data make it possible to generalize Eq.~(\ref{original}) to the case of the $n$-6 LJ systems and we report such a generalization here.     

\begin{table}
\caption{\label{Tab1} Coefficients ${\mathcal A}_{\rm L,S}$ and ${\mathcal B}_{\rm L,S}$ in Eq.~(\ref{MF}) for different exponents $n$ of the repulsive part of the $n$-6 LJ potential.  }
\begin{ruledtabular}
\begin{tabular}{lllllll}
$n$ & 12 & 11 & 10 & 9 & 8 & 7  \\ \hline
 ${\mathcal A}_{\rm S}$ & 1.865 & 1.929 & 2.036 & 2.230 & 2.656 & 4.054  \\
 ${\mathcal A}_{\rm L}$ & 2.166 & 2.181 & 2.243 & 2.403 & 2.811 & 4.225  \\
 ${\mathcal B}_{\rm S}$ & 0.972 & 1.039 & 1.158 & 1.374 & 1.832 & 3.271  \\
 ${\mathcal B}_{\rm L}$ & 0.581 & 0.707 & 0.881 & 1.154 & 1.674 & 3.199  \\
\end{tabular}
\end{ruledtabular}
\end{table}

\begin{figure*}
\includegraphics[width=14cm]{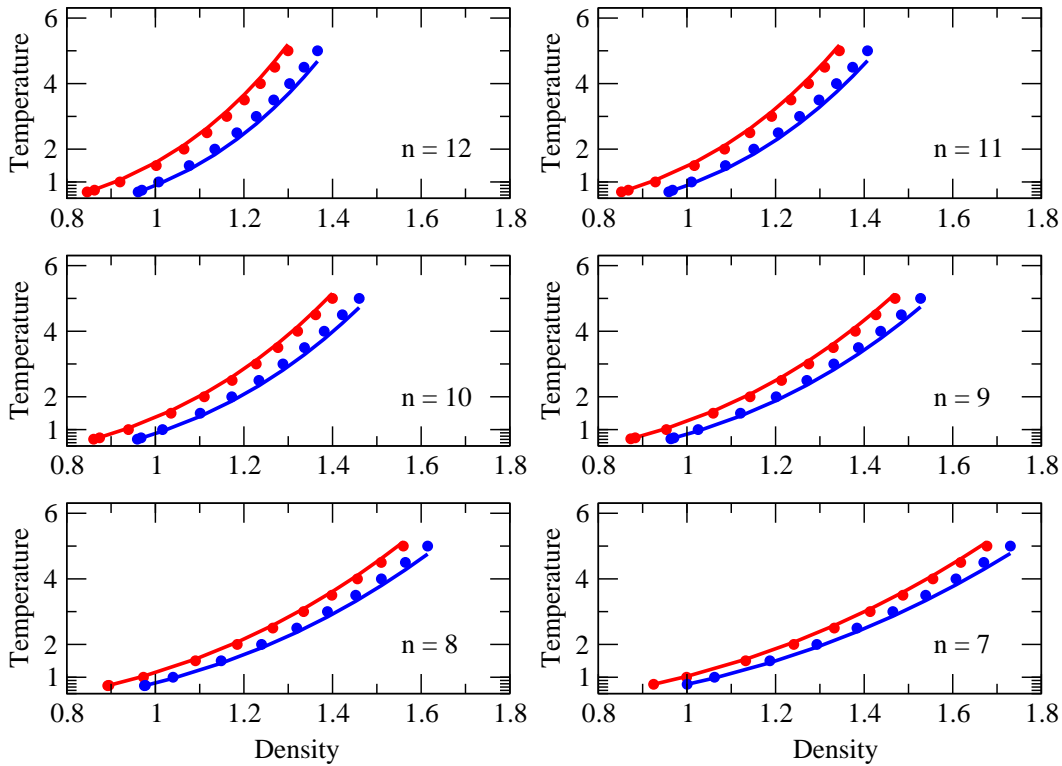}
\caption{Liquid-solid coexistence of the $n$-6 Lennard-Jones system ($n= 7-12$). Solid curves represent results from the analytical freezing and melting equations of this paper, where the red curves correspond to freezing and the blue curves correspond to melting. The solid circles represent the free energy calculation results obtained by Sousa et al.~\cite{Sousa2012}
} \label{Fig1}
\end{figure*}

We consider the generalized $n$-6 LJ potential of the form
\begin{equation}
U(r)= C_n\epsilon\left[(\sigma/r)^n - (\sigma/r)^6\right],  
\end{equation}
with
\begin{displaymath}
C_n=\left(\frac{n}{n-6}\right)\left(\frac{n}{6}\right)^{\frac{n}{n-6}}.
\end{displaymath}
The freezing and melting equations for these model interactions become
\begin{equation}\label{MF}
T_{\rm L,S}= {\mathcal A}_{\rm L,S}\rho^{\frac{n}{3}}-{\mathcal B}_{\rm L,S}\rho^2.
\end{equation}
The coefficients ${\mathcal A}$ have been determined from the fluid and solid densities at coexistence of the inverse-power-law systems (soft spheres) tabulated by Agrawal and Kofke.~\cite{Agrawal1995} The coefficients ${\mathcal B}$ have then been obtained using the triple point data of Ref.~\onlinecite{Sousa2012}. The results are summarized in Table~\ref{Tab1}. The resulting freezing and solid curves are plotted in Fig.~\ref{Fig1}, along with fluid-solid coexistence data of Sousa {\it et al}.~\cite{Sousa2012} Sufficiently good agreement is observed. Some differences can come from the densities of soft spheres at coexistence used to determine ${\mathcal A}$ values. For example, we have verified that for the special case $n=10$, using the more recent result $\rho_{\rm S}=1.327$ from Ref.~\onlinecite{Saija2006} (instead of $\rho_{\rm S}=1.338$ from Ref.~\onlinecite{Agrawal1995}) moves the melting line somewhat closer to the simulation results.   

In the high-temperature limit, Eq.~(\ref{MF}) does a very good job. To illustrate this we compare the fluid and solid densities at coexistence at $T=100$, tabulated in Table IV of Ref.~\onlinecite{Sousa2012} for the two special cases $n=12$ and $n=9$ with those obtained using Eq.~(\ref{MF}). For $n=12$ the precise simulation results are $\rho_{\rm L}=2.6342$ and $\rho_{\rm S}=2.7387$, while our results are $\rho_{\rm L}=2.6325$ and $\rho_{\rm S}=2.7547$. For $n=9$ the simulation results are $\rho_{\rm L}=3.6360$ and $\rho_{\rm S}=3.7404$, while our results are $\rho_{\rm L}=3.6332$ and $\rho_{\rm S}=3.7705$. The agreement for liquid densities is particularly good.     

Thus, the freezing and melting equations for $n$-6 LJ systems (\ref{MF}) appear rather accurate  in the high-density high-temperature limit and in the vicinity of the triple point. They also provide rather good interpolation between these limits.
Taking into account their simplicity and the absence of free parameters in the model (the coefficients ${\mathcal A}$ and ${\mathcal B}$ are {\it fixed} by the high-temperature asymptotes  and location of the triple points) the documented performance is more than satisfactory. After all, the phenomenological melting and freezing criteria leading to the functional form of Eq. (\ref{MF}) as well as the isomorph theory are all approximate.~\cite{Arxiv2016,KhrapakJCP2011,Heyes2015}  
More accurate equations (fits) can be of course constructed (for one recent examples see e.g. Ref.~\onlinecite{Heyes2015}), but these are normally less physically transparent and involve free parameters.   
 
To summarize, we have discussed very simple analytic equations for freezing and melting of the $n$-6 LJ system, which are reasonably accurate in the entire range of temperatures and densities.

This study was supported by the A*MIDEX grant (Nr.~ANR-11-IDEX-0001-02) funded by the French Government ``Investissements d'Avenir'' program.


\begin{thebibliography}{99}

\bibitem{KM2011} S. A. Khrapak and G. E. Morfill, J. Chem. Phys. {\bf 134}, 094108 (2011).
\bibitem{Rosenfeld1} Y. Rosenfeld, Chem. Phys. Lett. {\bf 38}, 591 (1976).
\bibitem{Rosenfeld2} Y. Rosenfeld, Mol. Phys. {\bf 32}, 963 (1976).
\bibitem{PRB2010} S. A. Khrapak, M. Chaudhuri, and G. E. Morfill, Phys. Rev. B {\bf 82}, 052101 (2010).

\bibitem{IngeJCP2012} T. S. Ingebrigtsen1, L. B{\o}hling, T. B. Schr{\o}der, and J. C. Dyre, J. Chem. Phys. {\bf 136}, 061102 (2012).
\bibitem{PRX2012} T. S. Ingebrigtsen, T. B. Schr{\o}der, and J. C. Dyre, Phys. Rev. X {\bf 2}, 011011 (2012).
\bibitem{DyrePRE2013} J. C. Dyre, Phys. Rev. E {\bf 87}, 022106 (2013).
\bibitem{Arxiv2016} L. Costigliola, T. B. Schr{\o}der, and J. C. Dyre, arXiv: 1602.03355v1 (2016).

\bibitem{Sousa2012} J. M. G. Sousa, A. L. Ferreira and M. A. Barroso, J. Chem. Phys. {\bf 136}, 174502 (2012).

\bibitem{Agrawal1995} R. Agrawal and D. A. Kofke, Mol. Phys. {\bf 85}, 23 (1995).

\bibitem{Saija2006} F. Saija, S. Prestipino, and P. V. Giaquinta, J. Chem. Phys. {\bf 124}, 244504 (2006).

\bibitem{KhrapakJCP2011} S. A. Khrapak, M. Chaudhuri, and G. E. Morfill, J. Chem. Phys. {\bf 134}, 054120 (2011).

\bibitem{Heyes2015} D. M. Heyes and A. C. Branka, J. Chem. Phys. {\bf 143}, 234504 (2015).

\end{thebibliography}
\end{document}